\documentclass[]{spie}  %>>> use for US letter paper
\usepackage{amsmath,amsfonts,amssymb}
\usepackage{graphicx}
\usepackage[colorlinks=true, allcolors=blue]{hyperref}

\usepackage{mathtools}
\usepackage{mathrsfs}
\makeatletter
\newcommand{\ostar}{\mathbin{\mathpalette\make@circled\star}}
\newcommand{\make@circled}[2]{%
  \ooalign{$\m@th#1\smallbigcirc{#1}$\cr\hidewidth$\m@th#1#2$\hidewidth\cr}%
}
\newcommand{\smallbigcirc}[1]{%
  \vcenter{\hbox{\scalebox{0.77778}{$\m@th#1\bigcirc$}}}%
}
\usepackage{array}
\usepackage{makecell}
\usepackage{multirow}
\newcolumntype{C}[1]{>{\centering\arraybackslash}p{#1}} % Centered column with cell aligned at the bottom
\newcolumntype{L}[1]{>{\raggedright\arraybackslash}m{#1}} % Left-flushed column with cell aligned at the middle
\newcolumntype{M}[1]{>{\centering\arraybackslash}m{#1}} % Centered column with text aligned at the middle
\newcolumntype{B}[1]{>{\centering\arraybackslash}b{#1}} % Centered column

\usepackage{graphicx}
\graphicspath{{./figures/}}

\usepackage{transparent}
\usepackage{xkeyval,xcolor}% http://ctan.org/pkg/{xkeyval,xcolor}
\makeatletter
\newlength{\sfp@hseplen}\newlength{\sfp@vseplen}
\define@cmdkey{subfigpos}[sfp@]{pos}[ul]{}% \sfp@pos
\define@cmdkey{subfigpos}[sfp@]{font}[\small]{}% \sfp@font
\define@cmdkey{subfigpos}[sfp@]{vsep}[0.75\baselineskip]{\setlength{\sfp@vseplen}{\sfp@vsep}}% \sfp@vsep
\define@cmdkey{subfigpos}[sfp@]{hsep}[3.5pt]{\setlength{\sfp@hseplen}{\sfp@hsep}}% \sfp@hsep
\newcommand{\subfigimg}[4][,]{%
        \setkeys{Gin,subfigpos}{pos,font,vsep,hsep,#1}% Set (default) keys
        \setbox1=\hbox{\includegraphics{#4}}% Store image in box
        \ifnum\pdfstrcmp{\sfp@pos}{ul}=0% UPPER LEFT placement of subfig label
                \leavevmode\rlap{\usebox1}% Print image
                \rlap{\hspace*{\sfp@hsep}\raisebox{\dimexpr\ht1-\sfp@vsep}{\transparent{#3}{\setlength{\fboxsep}{1pt}\colorbox{white}{%
\transparent{1}\sfp@font{#2}}}%
}}% Print label
                \phantom{\usebox1}% Insert appropriate spacing
        \else\ifnum\pdfstrcmp{\sfp@pos}{ur}=0% UPPER RIGHT placement of subfig label
                \leavevmode\usebox1% Print image
                \llap{\raisebox{\dimexpr\ht1-\sfp@vsep}{\sfp@font{#2}}\hspace*{\sfp@hsep}}% Print label
        \else\ifnum\pdfstrcmp{\sfp@pos}{lr}=0% LOWER RIGHT placement of subfig label
                \leavevmode\usebox1% Print image
                \llap{\raisebox{\sfp@vsep}{\sfp@font{#2}}\hspace*{\sfp@hsep}}% Print label
        \else% Assume LOWER LEFT placement of subfig label
                \leavevmode\rlap{\usebox1}% Print image
                \rlap{\hspace*{\sfp@hseplen}\raisebox{\sfp@vsep}{\sfp@font{#2}}}% Print label
                \phantom{\usebox1}% Insert appropriate spacing
        \fi\fi\fi
}
\newcommand{\fontfig}[1]{\small$\!\!$\color{#1}\textbf}
\newcommand{\AspectRatio}[1]{\dimexpr 1pt * \wd#1 / \ht#1 \relax} % Aspect ratio of a subfigure
\renewcommand{\refeq}[1]{Eq.~(\ref{#1})\xspace} % New definition from mathools...

\newcommand{\reffig}[1]{Fig.~\ref{#1}\xspace}
\newcommand{\refsub}[1]{#1} % How to refere to a subfigure.
\newcommand{\refsubfig}[2]{Fig.~\ref{#1}\refsub{#2}\xspace}
\newcommand{\refsubfigfull}[2]{Figure~\ref{#1}\refsub{#2}\xspace} % Full expansion of the name

\newcommand{\refpan}[1]{Panel~#1} % How to refere to a panel in a figure caption.
\newcommand{\refpans}[1]{Panels~#1} % How to refere to a panel in a figure caption.

\newcommand{\refsec}[1]{Sec.~\ref{#1}\xspace} % Section and subsection
\newcommand{\citet}[1]{Ref.~\citenum{#1}\xspace}
\newcommand{\refalg}[1]{Algorithm~\ref{#1}\xspace}
\newcommand{\reflin}[2]{Line~\ref{lin:#1:#2}\xspace} % Line of an algorithm
\usepackage{algorithm}
\usepackage[noend]{algpseudocode}
\newcommand{\commentalgo}[1]{\Comment{{#1}}} % for the comments in an algorithm
\DeclarePairedDelimiterX{\paren}[1]{(}{)}{#1}
\newcommand{\Paren}[1]{\paren*{#1}}
\let\brace=\undefined % Redifine \brace
\DeclarePairedDelimiterX{\brace}[1]{\{}{\}}{#1}
\newcommand{\Brace}[1]{\brace*{#1}}
\let\brack=\undefined % Redifine \brack
\DeclarePairedDelimiterX{\brack}[1]{[}{]}{#1}
\newcommand{\Brack}[1]{\brack*{#1}}
\DeclarePairedDelimiterX{\bbrack}[1]{\llbracket}{\rrbracket}{#1}%  [| ... |]

\DeclarePairedDelimiterX{\abs}[1]{\rvert}{\lvert}{#1}     %  | ... |

\DeclarePairedDelimiterX{\norm}[1]{\lVert}{\rVert}{#1}    % || ... ||

\DeclarePairedDelimiterX{\avg}[1]{\langle}{\rangle}{#1}   %  < ... >

\DeclarePairedDelimiterX{\ceil}[1]{\lceil}{\rceil}{#1}     % ceil operator

\DeclarePairedDelimiterX{\floor}[1]{\lfloor}{\rfloor}{#1}  % floor operator

\newcommand{\conv}{\star}				% Convolution symbol
\newcommand{\corr}{\ostar}				% Correlation symbol

\newcommand{\Tag}[1]{\text{#1}}			% Tag
\newcommand{\V}[1]{{\boldsymbol{#1}}}	% vector (with amsmath)
\newcommand{\data}{{d}}   		          	% data
\newcommand{\Vdata}{{\V{\data}}}          		
\newcommand{\obj}{{o}}   		          	% obj
\newcommand{\Vobj}{{\V{\obj}}}           	
\newcommand{\x}{{x}}   		          	 	% position
\newcommand{\Vx}{{\V{\x}}}            	  	
\newcommand{\psf}{{p}}            			% PSF
\newcommand{\Vpsf}{{\V{\psf}}}            	
\newcommand{\pmoon}{{\Vpsf^{\Tag{moon}}}}   	% PSF moon
\newcommand{\weight}{{w}}            		% Weight
\newcommand{\Vweight}{{\V{w}}}            	
\newcommand{\weightrob}{\weight^{\rob}}

\newcommand{\regpar}{{\mu}}            		% Regularisation parameter
\newcommand{\argmin}[2]{\underset{#1}{\text{argmin}}\;#2}

\newcommand{\dat}[2]{\mathscr{D}^{\Tag{#1}}\Paren{#2}}
\newcommand{\cost}[3]{\mathscr{C}^{\Tag{#1}}_{#2}\Paren{#3}}
\newcommand{\reg}[2]{\mathscr{R}^{\Tag{#1}}\Paren{#2}}
\newcommand{\rob}{\rho}

\newcommand{\throb}{\bar\weight^{\rob}}

\newcommand{\varRON}{v^{\Tag{ron}}}

\newcommand{\underseteq}[2]{\underset{\text{\refeq{#1}}}{#2}}
\usepackage[squaren,thickspace,thickqspace]{SIunits}
\newcommand{\percent}[1]{#1\,\%}     % percent
\usepackage{xspace}
\newcommand{\Kleopatra}{(216)~Kleopatra\xspace}
\newcommand{\Elektra}{(130)~Elektra\xspace}
\title{Blind and robust reconstruction of adaptive optics \\ point spread functions for asteroid deconvolution \\ and moon detection}

\author[a]{Anthony Berdeu}
\author[b]{Ferréol Soulez}
\author[c]{Kate Minker}
\author[c]{Benoit Carry}
\author[d]{Guillaume Bourdarot}
\author[b]{Antoine Kaszczyc}
\author[b]{Maud Langlois}

\affil[a]{LESIA, Observatoire de Paris, Université PSL, Sorbonne Université, Université Paris Cité, CNRS, 5 place Jules Janssen, 92195 Meudon, France}
\affil[b]{Univ Lyon, Univ Lyon1, ENS de Lyon, Centre de Recherche Astrophysique de Lyon, UMR 5574, F-69230, Saint-Genis-Laval, France}
\affil[c]{Université Côte d’Azur, Observatoire de la Côte d’Azur, CNRS, Laboratoire Lagrange, France}
\affil[d]{Max Planck Institute for extraterrestrial Physics, 85748 Garching, Germany}

\authorinfo{Send correspondence to Anthony Berdeu: \href{mailto:anthony.berdeu@obspm.fr}{anthony.berdeu@obspm.fr}}

\begin{document} 
\maketitle

\begin{abstract}
Initially designed to detect and characterize exoplanets, extreme adaptive optics systems (AO) open a new window on the solar system by resolving its small bodies. Nonetheless, despite the always increasing performances of AO systems, the correction is not perfect, degrading their image and producing a bright halo that can hide faint and close moons. Using a reference point spread function (PSF) is not always sufficient due to the random nature of the turbulence. In this work, we present our method to overcome this limitation. It blindly reconstructs the AO-PSF directly in the data of interest, without any prior on the instrument nor the asteroid's shape. This is done by first estimating the PSF core parameters under the assumption of a sharp-edge and flat object, allowing the image of the main body to be deconvolved. Then, the PSF faint extensions are reconstructed with a robust penalization optimization, discarding outliers on-the-fly such as cosmic rays, defective pixels and moons. This allows to properly model and remove the asteroid's halo. Finally, moons can be detected in the residuals, using the reconstructed PSF and the knowledge of the outliers learned with the robust method. We show that our method can be easily applied to different instruments (VLT/SPHERE, Keck/NIRC2), efficiently retrieving the features of AO-PSFs. Compared with state-of-the-art moon enhancement algorithms, moon signal is greatly improved and our robust detection method manages to discriminate faint moons from outliers.
\end{abstract}


\section{INTRODUCTION}
\label{sec:intro}

The study of asteroids witnessed a leap forward with the arrival of extreme adaptive optics (AO) systems that pushed further the performance in high-contrast and high-resolution imaging\cite{Vernazza:21_Large_program} of ground-based instruments. They are no longer limited by the atmospheric turbulence that corrugates the incident wavefront of the observed target\cite{Roddier:81}, with AO systems now commonly deployed in observatories\cite{Tyson:15_principles_of_AO, Jovanovic:15_XAO}. In this context, the study of the point spread function (PSF) after an AO system correction emerged in order to improve the data post-processing via model-fitting or reconstruction techniques\cite{Beltramo:20_PSF_reconstruction_Review}. For asteroid image recovery, a fine knowledge of the PSFs is needed to avoid strong deconvolution artifacts\cite{Marchis:06_moon_detection, Fetick:19_Vesta, Fetick:20_param_marignal, Lau:23_prior_AOPSF}. But due to the random nature of AO-PSFs, the direct estimation of the PSF parameters from the AO telemetry\cite{Veran:97_PSF_AO_telemetry, Clenet:08_NACO_PSF_recons} or reference PSFs obtained on calibration sources (internal or natural stars) before or after the observation\cite{Mugnier:04_Mistral} are not always sufficient.

The only solution is thus to extract and reconstruct the PSF directly from the data of interest, a problem known as blind deconvolution\cite{Stockham:75_blind_deconvolution, Thiebaut:95_blind_deconvolution, Soulez:12, Fetick:20_param_marignal}. Marginal approaches\cite{Blanc:03_Marginal_Zernike, Beltramo:20_PSF_reconstruction_Review}, that split the contributions of the PSF and the object in the problem, have proved to be efficient when combined with parametric PSF models that strongly limit the number of unknowns to fit\cite{Fetick:19_model_based_AOPSF}. But these simplified methods cannot fully grasp the complexity of real and potentially broadband AO-PSFs. For Solar System bodies, this approximate knowledge of the PSF limits the study of their close vicinity and the detection of faint companions\cite{Showalter:06_Uranus, Assafin:08_digital_coronography, Vernazza:21_Large_program}, buried in the bright halo induced by the PSF faint extensions. Techniques adapted from exoplanet detection algorithms, based on local averaging or median filters\cite{Marchis:06_moon_detection, Assafin:08_digital_coronography, Pajuelo:18_Carry_coronagraph} are not always sufficient\cite{Yang:16_Elektra_Minerva}.

More versatile methods than parametric approaches\cite{Berdeu:22_Elektra} are consequently needed with the challenge to recover the AO-PSF extensions, several orders of magnitude fainter than the PSF core. This paper is a continuation of a method developed to jointly recover the 2D deconvolved images of the object and the PSF while being robust to outliers, see \citet{Berdeu:24_deconv}. This method provides a more general and blind approach for more complex AO-PSFs with limited priors on the object and the PSF. In this approach, outliers are defective pixels or pixels hit by cosmic rays, as well as potential companion orbiting the main body. In this work, after a brief reminder of the method, see \refsec{sec:deconv}, we focus on the possibility to detect faint companions in the residuals after the removal from the data of the bright halo, see \refsec{sec:results}.

\section{OVERVIEW OF THE PROPOSED METHOD ON SIMULATIONS}

\subsection{Blind Deconvolution and Halo Removal}
\label{sec:deconv}

\begin{figure}[p!] % fig:overview
        \centering
        
        % Internal command of the figure for the automatic sizing
        % Path of the files
        \newcommand{\PathFig}{Fig_overview/}
        
        % Line ratio
        \newcommand{\LineRatio}{0.99}
        
        % Font of the text in the figure
        \newcommand{\fontTxt}[1]{\textbf{\scriptsize #1}}
        
        % Width of the text boxes        
        \newcommand{\widthTxt}{12pt}
        
        % Width of the text boxes        
        \newcommand{\sizeTxt}[1]{\large{#1}}
        
        % Vertical space between lines
        \newcommand{\spaceLine}{1.6cm}
        
        % First line
        \newcommand{\widthFig}{\dimexpr (\linewidth - \widthTxt * 4)}
        
        \newcommand{\subfigColor}{white}        
        
        % Getting the size of the boxes
        \sbox1{\includegraphics{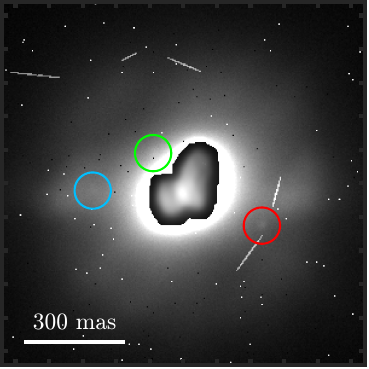}}
        \sbox2{\includegraphics{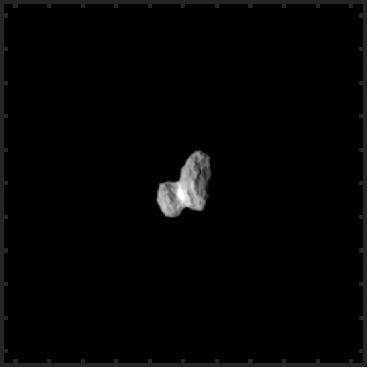}}
        \sbox3{\includegraphics{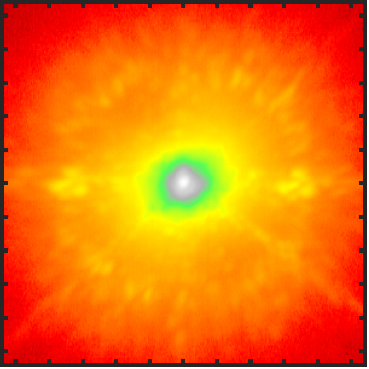}}
        \sbox4{\includegraphics{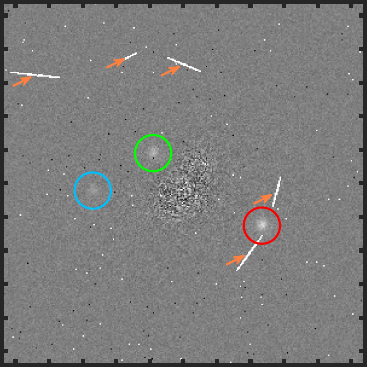}}
        \sbox5{\includegraphics{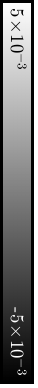}}
        
        % Defining column width command
        \newcommand{\ColumnWidth}[1]
                {\dimexpr \LineRatio \widthFig * \AspectRatio{#1} / (\AspectRatio{1} + \AspectRatio{2} + \AspectRatio{3} + \AspectRatio{4} + \AspectRatio{5}) \relax
                }
        \newcommand{\ColumnGap}{\hspace {\dimexpr \widthFig /15 - \LineRatio\widthFig /15 }}

        % Figure table
        \begin{tabular}{
                @{\ColumnGap}
                M{\widthTxt}
                @{\ColumnGap}
                M{\ColumnWidth{1}}
                @{\ColumnGap}
                M{\widthTxt}
                @{\ColumnGap}
                M{\ColumnWidth{2}}
                @{\ColumnGap}
                M{\widthTxt}
                @{\ColumnGap}
                M{\ColumnWidth{3}}
                @{\ColumnGap}
                M{\widthTxt}
                @{\ColumnGap}
                M{\ColumnWidth{4}}
                @{\ColumnGap\ColumnGap\ColumnGap\ColumnGap\ColumnGap\ColumnGap}
                M{\ColumnWidth{5}}
                @{\ColumnGap}
                }
                
                % Title line
                &
                \fontTxt{Noisy data / Model}&
                &
                \fontTxt{Object}&
                &
                \fontTxt{Point spread function}&
                &
                \fontTxt{Noise / Residuals}&
                \\                
                
                % First line
                \rotatebox[origin=l]{90}{\fontTxt{Forward model}} &
                \subfigimg[width=\linewidth,pos=ul,font=\fontfig{\subfigColor}]{$\;$(a1)}{0.0}{\PathFig Data_sim.pdf} &
                \sizeTxt{$=$} &
                \subfigimg[width=\linewidth,pos=ul,font=\fontfig{\subfigColor}]{$\;$(a2)}{0.0}{\PathFig Obj_sim.pdf} &
                \sizeTxt{$\conv$} &
                \subfigimg[width=\linewidth,pos=ul,font=\fontfig{\subfigColor}]{$\;$(a3)}{0.0}{\PathFig PSF_sim.pdf} &
                \sizeTxt{$+$} &
                \subfigimg[width=\linewidth,pos=ul,font=\fontfig{\subfigColor}]{$\;$(a4)}{0.0}{\PathFig Res_sim.pdf} &
                \subfigimg[width=\linewidth,pos=ul,font=\fontfig{\subfigColor}]{}{0.0}{\PathFig Res_bar.pdf}
                \\[\spaceLine]             
                
                % Second line
                \rotatebox[origin=l]{90}{\fontTxt{1. PSF core estimate}} &
                \subfigimg[width=\linewidth,pos=ul,font=\fontfig{\subfigColor}]{$\;$(b1)}{0.0}{\PathFig Data_core.pdf} &
                \sizeTxt{$=$} &
                \subfigimg[width=\linewidth,pos=ul,font=\fontfig{\subfigColor}]{$\;$(b2)}{0.0}{\PathFig Obj_core.pdf} &
                \sizeTxt{$\conv$} &
                \subfigimg[width=\linewidth,pos=ul,font=\fontfig{\subfigColor}]{$\;$(b3)}{0.0}{\PathFig PSF_core.pdf} &
                \sizeTxt{$+$} &
                \subfigimg[width=\linewidth,pos=ul,font=\fontfig{\subfigColor}]{$\;$(b4)}{0.0}{\PathFig Res_core} &
                \subfigimg[width=\linewidth,pos=ul,font=\fontfig{\subfigColor}]{}{0.0}{\PathFig Res_core_bar.pdf}
                \\[\spaceLine]             
                
                % Third line
                \rotatebox[origin=l]{90}{\fontTxt{2. Object deconvolution}} &
                \subfigimg[width=\linewidth,pos=ul,font=\fontfig{\subfigColor}]{$\;$(c1)}{0.0}{\PathFig Data_deconv.pdf} &
                \sizeTxt{$=$} &
                \subfigimg[width=\linewidth,pos=ul,font=\fontfig{\subfigColor}]{$\;$(c2)}{0.0}{\PathFig Obj_deconv.pdf} &
                \sizeTxt{$\conv$} &
                \subfigimg[width=\linewidth,pos=ul,font=\fontfig{\subfigColor}]{$\;$(c3)}{0.0}{\PathFig PSF_deconv.pdf} &
                \sizeTxt{$\Rightarrow$} &
                \subfigimg[width=\linewidth,pos=ul,font=\fontfig{\subfigColor}]{$\;$(c4)}{0.0}{\PathFig Res_deconv.pdf} &
                \subfigimg[width=\linewidth,pos=ul,font=\fontfig{\subfigColor}]{}{0.0}{\PathFig Res_core_bar.pdf}
                \\[\spaceLine]             
                
                % Fourth line
                \rotatebox[origin=l]{90}{\fontTxt{3. PSF deconvolution}} &
                \subfigimg[width=\linewidth,pos=ul,font=\fontfig{\subfigColor}]{$\;$(d1)}{0.0}{\PathFig Data_wing.pdf} &
                \sizeTxt{$=$} &
                \subfigimg[width=\linewidth,pos=ul,font=\fontfig{\subfigColor}]{$\;$(d2)}{0.0}{\PathFig Obj_wing.pdf} &
                \sizeTxt{$\conv$} &
                \subfigimg[width=\linewidth,pos=ul,font=\fontfig{\subfigColor}]{$\;$(d3)}{0.0}{\PathFig PSF_wing.pdf} &
                \sizeTxt{$\Rightarrow$} &
                \subfigimg[width=\linewidth,pos=ul,font=\fontfig{\subfigColor}]{$\;$(d4)}{0.0}{\PathFig Res_wing.pdf} &
                \subfigimg[width=\linewidth,pos=ul,font=\fontfig{\subfigColor}]{}{0.0}{\PathFig Res_bar.pdf}
                \\[-2pt] 
                
                % Color bar line
                &
                \subfigimg[width=\linewidth,pos=ul,font=\fontfig{\subfigColor}]{}{0.0}{\PathFig Data_bar.pdf} &
                &
                \subfigimg[width=\linewidth,pos=ul,font=\fontfig{\subfigColor}]{}{0.0}{\PathFig Obj_bar.pdf} &
                &
                \subfigimg[width=\linewidth,pos=ul,font=\fontfig{\subfigColor}]{}{0.0}{\PathFig PSF_bar.pdf} &
                &
                &
                \\

        \end{tabular}        
        \caption{\label{fig:overview} Overview of the deconvolution method on a simulation based on a photo of 67P/Churyumov–Gerasimenko by Rosetta (European Space Agency) with three synthetic moons. \textit{First line~(a)}~--~Forward model: the noisy data~$\Vdata$~(a1) is the convolution of an extended object~$\Vobj$~(a2) with the AO-corrected PSF of the instrument~$\Vpsf$~(a3) plus a nuisance term~$\V{n}$~(a4). This noise is composed of the acquisition noises (readout + photon), defective pixels (``pepper \&{} salt'' pattern), cosmic ray impacts (orange arrows) and signal from potential moons orbiting the main object (coloured circles). \textit{Second line~(b)}~--~Step~1. The data is approximated by the convolution of a binary object~(b2, and orange edges in b1) with a parametric PSF core~(b3). \refpan{b4}: Residuals of the fit performed on the pixels of \refpan{b1} not shaded in red. \textit{Third line~(c)}~--~Step~2. The estimated PSF core~(c3=b3) is used to deconvolve the object~(c2). \refpan{c1}: Model of the convolution. \refpan{c4}: Residuals. \textit{Fourth line (d)}~--~Step~3. The segmented object~(d2) is used to deconvolve the PSF wings~(d3). \refpan{d1}: Model of the convolution. \refpan{d4}: Residuals. \textit{First~(1) and second~(2) columns}~--~To emphasize both the main body and the surrounding halo a dual linear scale is used to insert the main body in its halo, as noted by the ``/'' in the color bars. \textit{Third~(3) column}~--~The PSF are normalized to peak at one for the display.}
\end{figure}

The overview of the method for blind deconvolution and halo removal, applied on a simulation, is given in \reffig{fig:overview}. This simulation is based on (I) a photo of 67P/Churyumov–Gerasimenko by Rosetta (European Space Agency) as a reference, scaled to the resolution of a standard asteroid observation of a few hundred milliarcseconds, see \reffig{fig:overview}{a2}, and (II) a PSF  obtained on a star with the Zurich IMaging POLarimeter instrument\cite{Schmid:18_ZIMPOL} (ZIMPOL), see \refsubfig{fig:overview}{a3}. Three synthetic moons are injected, highlighted by the colored circles.

The forward model of the problem canonically states that the image data,~$\Vdata$, is the convolution~$\conv$ of the extended object,~$\Vobj$, with the long exposure PSF,~$\Vpsf$, that combines the telescope and instrument response and the AO residuals\cite{Fetick:20_param_marignal},
\begin{equation}
	\label{eq:forward_model}
	\Vdata = \Paren{\Vobj \conv \Vpsf} + \V{n}
	\,,
\end{equation}
where~$\V{n}$ is a nuisance term. This equation is pictured in \reffig{fig:overview}{a}. As is seen in \refsubfig{fig:overview}{a4}, the nuisance term encompasses the classical noise terms (detector readout and photon shot noises) as well as outliers such as defective pixels, cosmic rays  (orange arrows) and potential moons (colored circles) that cannot be explained by the convolution of the PSF with an extended object. The objectives of the blind deconvolution method are to split~$\Vobj$ and~$\Vpsf$ only from the knowledge of~$\Vdata$ without being corrupted by~$\V{n}$ and remove their contributions from the data to enhance moon signal.

The PSF presented in \refsubfig{fig:overview}{a3} emphasizes all the complexity of an AO-PSFs. The faint extensions are responsible for the extended and structured bright halo visible in \refsubfig{fig:overview}{a1} in which the moon are hidden.

As further described in details in \citet{Berdeu:24_deconv}, \refeq{eq:forward_model} is inverted by minimizing the regularized cost function
\begin{equation}
	\label{eq:inverse_problem}
	\Paren{\tilde{\Vobj}, \tilde{\Vpsf}} \gets \argmin{\Vobj\geq\V{0}, \Vpsf\geq\V{0}}{\Brace{
	\cost{}{}{\Vobj, \Vpsf} =
	\dat{wls}{\Vdata, \Vobj \conv \Vpsf ,\Vweight}
	+ 
	\regpar^{\Tag{obj}}\reg{obj}{\Vobj}
	+ 
	\regpar^{\Tag{psf}}\reg{psf}{\Vpsf}
	}}
	\,,
\end{equation}
where, (I) $\mathscr{D}^{\Tag{wls}}$ is defined as the weighted least square difference (wls):
\begin{equation}
	\label{eq:WLS}
	\dat{wls}{\V{\varphi}_{1},\V{\varphi}_{2},\Vweight}
	{}\triangleq{}
	\frac{1}{2}\sum_{\Vx} \weight\Paren{\Vx}
	\Paren{\varphi_{1}\Paren{\Vx} -\varphi_{2}\Paren{\Vx}}^{2}
	\,,
\end{equation}
(II) $\reg{obj}{\Vobj}$ is a regularization favoring smooth objects with sharp edges, by encouraging the sparsity of spatial gradients\cite{Rudin:92_TV, Charbonnier:97_TV} as classically used in asteroid deconvolution\cite{Mugnier:04_Mistral, Fetick:20_param_marignal, Berdeu:22_Elektra, Lau:23_prior_AOPSF, Yan:23_myopic_MCMC}:
\begin{equation} % eq:reg_obj
	\label{eq:reg_obj}
	\reg{obj}{\V{\varphi}}
	{}\triangleq{}
	\sum_{\x}\Brack{\sqrt{\Paren{\brack{\nabla_{1}\varphi\Paren{\x}}^{2}+\brack{\nabla_{2}\varphi\Paren{\x}}^{2}+\brack{\epsilon^{\Tag{obj}}}^{2}}}-\epsilon^{\Tag{obj}}}
	\,,
\end{equation}
and (III) $\reg{psf}{\Vpsf}$ consists of the classical $\ell^2$-norm on the gradient but applied to the logarithm of the PSF:
\begin{equation} % eq:reg_psf
	\label{eq:reg_psf}
	\reg{psf}{\V{\varphi}}
	{}\triangleq{}
	\sum_{\x}\Brack{\nabla_{1}\Brack{\ln\Paren{\varphi\Paren{\x}}}}^{2}+\Brack{\nabla_{2}\Brack{\ln\Paren{\varphi\Paren{\x}}}}^{2}
	\,.
\end{equation}
In the data fidelity term of \refeq{eq:WLS}, the weight term $\Vweight$ is the inverse of the data variance to whiten the residuals~$\Vdata - \Vdata^{\Tag{mod}}$ with
\begin{equation}
	\label{eq:data_mod}
	\Vdata^{\Tag{mod}} = \tilde{\Vobj} \conv \tilde{\Vpsf}
	\,.
\end{equation}
The weight term $\Vweight$ is given by\cite{Mugnier:04_Mistral, Fetick:19_model_based_AOPSF, Berdeu:20_PIC}:
\begin{equation}
	\label{eq:weight}
	\weight\Paren{\Vx} = 1/\Paren{\eta\data\Paren{\Vx}+\varRON}
	\,,
\end{equation}
where~$\varRON$ is the readout noise and~$\eta$ the factor of the photon noise to convert the intensity into a variance. In \refeq{eq:reg_obj}, $\nabla_{1}$ and $\nabla_{2}$ correspond to finite difference operators along the two spatial dimensions of the image and $\epsilon^{\Tag{obj}}>0$ is a threshold controlling the transition between a $\ell^1$-norm (edge-preserving) and a $\ell^2$-norm (smoothness). $\epsilon^{\Tag{obj}}$ also ensures that \refeq{eq:reg_obj} is differentiable at zero. In \refeq{eq:reg_psf}, the norm on the 2D-gradient ensures a smooth reconstruction, while the logarithm acts as a ``dynamic whitening'' term on the PSF wings of $\Vpsf$ that are multiple orders of magnitude fainter than its core. Finally, in \refeq{eq:inverse_problem}, $\regpar^{\Tag{obj}}$ and $\regpar^{\Tag{psf}}$ are hyperparameters to balance the regularizations on the object and the PSF compared to the data fidelity term. We briefly introduce hereafter the steps of the method described in \citet{Berdeu:24_deconv} to solve the minimzation problem of \refeq{eq:inverse_problem}.

(I) Firstly, as is pictured in \refsubfig{fig:overview}{b}, the initialization is obtained by solving an approximation of the problem. In a crude simplification, the data of \refsubfig{fig:overview}{b1} can been seen as a sharp-edged flat object, \refsubfig{fig:overview}{b2}, convolved with a simple PSF core, \refsubfig{fig:overview}{b3}. In this step, the estimated object~$\tilde{\Vobj}$ is obtained by simply applying a threshold on the blurred data and a parametric model is used for the PSF~$\tilde{\Vpsf}$ (2D Moffat pattern\cite{Moffat:69}). To avoid any corruption of the fit by the halo extensions, the pixels shaded in red in \refsubfig{fig:overview}{b1} are removed from the fit. As seen in the residuals of \refsubfig{fig:overview}{b4}, most of the signal of the bright core of the halo is removed with this approach, but the faint extensions of the halo still hide the moons to recover.

(II) Secondly, as is pictured in \refsubfig{fig:overview}{c}, this PSF core is used to deconvolve the main extended object~$\tilde{\Vobj}$, \refsubfig{fig:overview}{c2}, fixing~$\tilde{\Vpsf}$, \refsubfig{fig:overview}{c3}. The physical hard constraint $\Vobj\geq\V{0}$ imposes a positive object and limits oscillating artifacts close to its sharp edges, as commonly seen in deconvolution problems\cite{Marchis:06_moon_detection, Fetick:20_param_marignal}.

\begin{figure}[t!] % fig:weight
        \centering
                
        % Line ratio
        \newcommand{\LineRatio}{0.75}
        \newcommand{\widthBar}{3pt}
        \newcommand{\widthFig}{\dimexpr (\linewidth - \widthBar * 3)}

        % Vertical space between lines
        \newcommand{\spaceLine}{1.8cm}
        
        % Font of the text in the figure
        \newcommand{\fontTxt}[1]{\textbf{\small #1}}
        
        \newcommand{\subfigColor}{black}        
        
        % Getting the size of the boxes
        \newcommand{\FigOne}{Fig_weight/Weight_map}  
        \newcommand{\FigTwo}{Fig_weight/Weight_bar}  
        \sbox1{\includegraphics{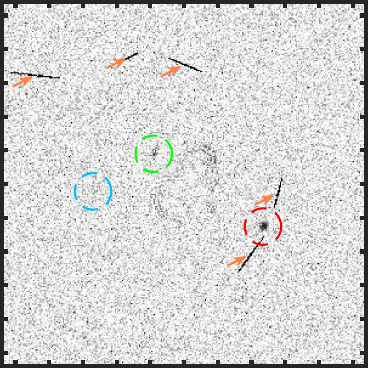}}
        \sbox2{\includegraphics{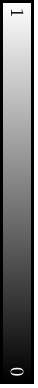}}
        
        % Defining column width command
        \newcommand{\ColumnWidth}[1]
                {\dimexpr \LineRatio \widthFig * \AspectRatio{#1} / (3*\AspectRatio{1} + 3*\AspectRatio{2}) \relax
                }
        \newcommand{\ColumnGap}{\hspace {\dimexpr \widthFig /10 - \LineRatio\widthFig /10 }}

        % Figure table
        \begin{tabular}{
                @{\ColumnGap}
                M{\ColumnWidth{1}}
                @{\hspace{\widthBar}}
                M{\ColumnWidth{2}}
                @{\ColumnGap}
                M{\ColumnWidth{1}}
                @{\hspace{\widthBar}}
                M{\ColumnWidth{2}}
                @{\ColumnGap}
                M{\ColumnWidth{1}}
                @{\hspace{\widthBar}}
                M{\ColumnWidth{2}}
                @{\ColumnGap}
                }

                % First line
                \subfigimg[width=\linewidth,pos=ul,font=\fontfig{white}]{$\;$(a)}{0.0}{Fig_detection/Res_map} &
                \subfigimg[width=\linewidth,pos=ul,font=\fontfig{\subfigColor}]{}{0.0}{Fig_detection/Res_bar} &
                \subfigimg[width=\linewidth,pos=ul,font=\fontfig{\subfigColor}]{$\;$(b)}{0.0}{Fig_weight/Weight_map} &
                \subfigimg[width=\linewidth,pos=ul,font=\fontfig{\subfigColor}]{}{0.0}{Fig_weight/Weight_bar} &
                \subfigimg[width=\linewidth,pos=ul,font=\fontfig{\subfigColor}]{$\;$(c)}{0.0}{Fig_weight/Valid_map} &
                \subfigimg[width=\linewidth,pos=ul,font=\fontfig{\subfigColor}]{}{0.0}{Fig_weight/Weight_bar}
                \\
                
        \end{tabular}        
        \caption{\label{fig:weight} Automatic removal of outliers with respect to the model. \refpan{a}: Residuals of the halo removal~$\Vdata - \Vdata^{\Tag{mod}}$, for information, the deconvolved main body image was inserted with the same dynamics as in \refsubfig{fig:overview}{d2}. \refpan{b}: Equivalent weights of the robust penalisation,~$\weightrob\paren{\sqrt{\Vweight}\paren{\Vdata - \Vdata^{\Tag{mod}}}}$. \refpan{c}: Discarding of pixels (black) below a threshold,~$\throb$.
        }
\end{figure}

(III) Thirdly, as is pictured in \refsubfig{fig:overview}{d}, the deconvolution paradigm is reversed and the object is used to deconvolve the faint PSF extensions~$\tilde{\Vpsf}$, so-called wings, \refsubfig{fig:overview}{d3}, fixing~$\tilde{\Vobj}$, \refsubfig{fig:overview}{d2} with the physical positivity constraint $\Vpsf\geq\V{0}$. In this context, outliers must be carefully handled to avoid any corruption of the model and identified on the fly. A robust penalization approach has been implemented\cite{Zoubir:18_robust_stat,Flasseur:19_PhD}, where the conventional quadratic penalization of the problem is replaced by a robust estimator. This estimator is approximately quadratic around zero but grows sub-quadratically for large deviations to reduce their impact\cite{Hogg:79, Huber:96}. An iterative reweighted least squares\cite{Holland:77, Sigl:16_nonlinear_IRLS} (IRLS) method is used: a sequence of least squares problems are solved with their weights iteratively updated with a robust estimator. The robust estimator~$\rob$ on the residuals~$\Vdata - \Vdata^{\Tag{mod}}$ is the Cauchy cost function\cite{Holland:77}:
\begin{equation}
	\rob\Paren{r}
	{}\triangleq{}
	\frac{\gamma^2}{2}\ln\Paren{1+ r^2 / \gamma^2}
	\text{ with }
	\gamma = 2.385
	\,.
\end{equation}
The robust weights
\begin{equation} % eq:Cauchy_weight
	\label{eq:Cauchy_weight}
	\weightrob\Paren{r}=\frac{1}{r}\frac{\partial\rob\Paren{r}}{\partial r}=\Paren{1+r^2 / \gamma^2}^{-1}
	\,,
\end{equation}
is the correction factor to apply to the weights in the least square error of \refeq{eq:WLS} in the context of IRLS. \refsubfigfull{fig:weight}{b} gives an example of a robust weight map, obtained by applying $\weightrob$ of \refeq{eq:Cauchy_weight} to a residual map~$\Vdata - \Vdata^{\Tag{mod}}$, \refsubfig{fig:weight}{a}. The outliers can directly be identified in black, with equivalent weights tending towards zero. It is then possible to discard them on the fly, using a conservative threshold ~$\throb=\percent{50}$ as is shown in \refsubfig{fig:weight}{c}. This is done by updating \refeq{eq:weight} while accounting for the data model $\data^{\Tag{mod}}$:
\begin{equation}
	\label{eq:weight_mod}
	\weight\Paren{\Vx} =
	\begin{cases}
		0 \text{ if } \weightrob\Paren{\Vx} \leq \throb
		\\
		1/\Paren{\eta\data^{\Tag{mod}}\Paren{\Vx}+\varRON} \text{ otherwise}
	\end{cases}
	\,,
\end{equation}
preventing any further corruption of the deconvolved PSF wings by strong outliers.

Now that a better model of the PSF is known, especially around its core, it is possible to re-estimate the object. Alternating the steps (II) and (III) while updating the map of discarded outliers allows to refine the models of the object and the PSF. Finally, after convergence, the halo model, \refsubfig{fig:overview}{d1}, is clean: it is not polluted by cosmic ray or moon signals. It can be removed and the moons can be seen in the residuals, \refsubfig{fig:overview}{d4}.

\subsection{Robust Moon Detection}
\label{sec:detect}

\begin{figure}[h!]
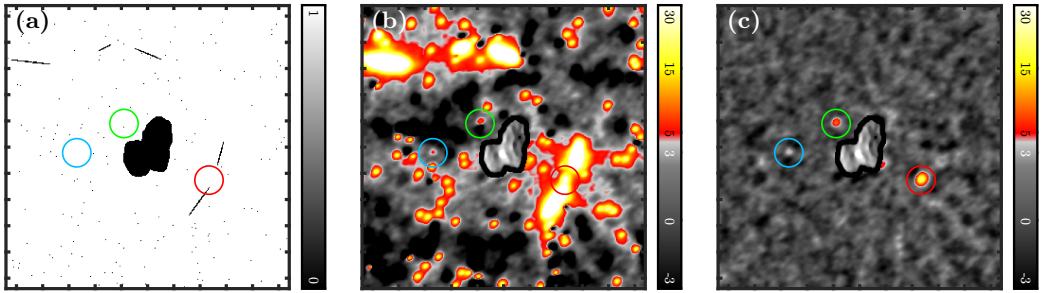
 % fig:detection
        \centering
                
        % Internal command of the figure for the automatic sizing
        % Path of the files
        \newcommand{\PathFig}{Fig_detection/}
        
        % Line ratio
        \newcommand{\LineRatio}{0.75}
        \newcommand{\widthBar}{3pt}
        \newcommand{\widthFig}{\dimexpr (\linewidth - \widthBar * 3)}

        % Vertical space between lines
        \newcommand{\spaceLine}{1.8cm}
        
        % Font of the text in the figure
        \newcommand{\fontTxt}[1]{\textbf{\small #1}}
        
        \newcommand{\subfigColor}{white}        
        
        % Getting the size of the boxes
        \newcommand{\FigOne}{\PathFig Valid_map.pdf}  
        \newcommand{\FigTwo}{\PathFig Weight_bar.pdf}  
        \sbox1{\includegraphics{\FigOne}}
        \sbox2{\includegraphics{\FigTwo}}
        
        % Defining column width command
        \newcommand{\ColumnWidth}[1]
                {\dimexpr \LineRatio \widthFig * \AspectRatio{#1} / (3*\AspectRatio{1} + 3*\AspectRatio{2}) \relax
                }
        \newcommand{\ColumnGap}{\hspace {\dimexpr \widthFig /10 - \LineRatio\widthFig /10 }}

        % Figure table
        \begin{tabular}{
                @{\ColumnGap}
                M{\ColumnWidth{1}}
                @{\hspace{\widthBar}}
                M{\ColumnWidth{2}}
                @{\ColumnGap}
                M{\ColumnWidth{1}}
                @{\hspace{\widthBar}}
                M{\ColumnWidth{2}}
                @{\ColumnGap}
                M{\ColumnWidth{1}}
                @{\hspace{\widthBar}}
                M{\ColumnWidth{2}}
                @{\ColumnGap}
                }

                % First line
                \subfigimg[width=\linewidth,pos=ul,font=\fontfig{black}]{$\;$(a)}{0.0}{\PathFig Valid_map.pdf} &
                \subfigimg[width=\linewidth,pos=ul,font=\fontfig{\subfigColor}]{}{0.0}{\PathFig Weight_bar.pdf} &
                \subfigimg[width=\linewidth,pos=ul,font=\fontfig{\subfigColor}]{$\;$(b)}{0.0}{\PathFig sig_map.pdf} &
                \subfigimg[width=\linewidth,pos=ul,font=\fontfig{\subfigColor}]{}{0.0}{\PathFig Sig_bar.pdf} &
                \subfigimg[width=\linewidth,pos=ul,font=\fontfig{\subfigColor}]{$\;$(c)}{0.0}{\PathFig sig_map_BP.pdf} &
                \subfigimg[width=\linewidth,pos=ul,font=\fontfig{\subfigColor}]{}{0.0}{\PathFig Sig_bar.pdf}
                \\
                
        \end{tabular}        
        \caption{\label{fig:detection} Detection of potential moons. \refpan{a}: Discarding main body vicinity and isolated pixels below a threshold (black). \refpans{b,c}: Significance map~$\V{\sigma}$ of moon detection in \refsubfig{fig:weight}{a} directly on the residuals map~(\refpan{b}) and accounting for the discarded pixels~(\refpan{c}). \refpans{b,c}: For information, the deconvolved main body image was inserted with the same dynamics as in \refsubfig{fig:overview}{d2}.}
\end{figure}

Now that a residual map cleaned from the halo pollution has been obtained, the objective of this section is to provide a detection map of potential moons orbiting the main object. This is done by estimating the likelihood to find a moon signal~$\pmoon$, in the halo residuals~$\Vdata - \Vdata^{\Tag{mod}}$ of \refsubfig{fig:detection}{a}. Since moons are not resolved, they behave as point sources and it naturally comes that $\pmoon\propto\tilde{\Vpsf}$. As explained in \citet{Thiebaut:16_detection}, this likelihood is given by the normalized correlation between~$\pmoon$ and these residuals, weighted by~$\Vweight$,
\begin{equation} % eq:sig_map
	\label{eq:sig_map}
	\V{\sigma} = \frac{\pmoon\corr\Brack{\Vweight\Paren{\Vdata - \Vdata^{\Tag{mod}}}}}{\sqrt{\Paren{\brack{\pmoon}^{2}\corr\Vweight}}}
	\,,
\end{equation}
where $\corr$ denotes for the correlation symbol.

In principle, as in \refeq{eq:weight_mod}, $\weight\Paren{\Vx}$ is given by the model in the absence of moon: $1/\Paren{\eta\data^{\Tag{mod}}\Paren{\Vx}+\varRON}$. A moon would thus give a significant signal compared to this awaited noise level. Nonetheless, doing so leads to the significance map of \refsubfig{fig:detection}{b}. This map is highly corrupted by other strong outliers such as cosmic rays or hot pixels. Moons cannot be disentangled from these artifacts.

To clean this map, it is possible to further exploit the robust weight map of \refsubfig{fig:weight}{b} obtained during the PSF wing deconvolution in \refsec{sec:deconv}. Indeed, it contains the information on possible strong outliers. The technique to generate the robust significance maps is given in \refalg{alg:sig}. Applying \reflin{sig}{th} to discard outliers on-the-fly implies to find the isolated pixels. They are identified by dilating and then eroding by one pixel the binary map of the pixels above a \percent{10} threshold\cite{Gonzalez:20_Matlab} (morphological operations). The pixels which disappear in the process are considered as isolated. In practice the pixels in the vicinity of the main body below the threshold of \percent{10} are also discarded to avoid any bias coming from the bad fit close to the object's edges, as shown in \refsubfig{fig:detection}{a}. This empirical value of \percent{10} only excludes strong outliers. Higher values may tag pixel clusters that would not be considered as isolated and thus not discarded. As mentioned in \citet{Thiebaut:16_detection}, the maps are zero-padded by a factor of two to avoid aliasing when computing the correlations of \reflin{sig}{map}.

\begin{algorithm}
\caption{\label{alg:sig} Moon significance map.}
\begin{algorithmic}[1]
\small
	\State $\pmoon \gets \tilde{\Vpsf}$
		\commentalgo{Moon PSF}

	\State $\pmoon \gets \pmoon/\sum_{\Vx}{\psf^{\Tag{moon}}\Paren{\Vx}}$
		\commentalgo{Normalization of the PSF}

	\State $\weight\Paren{\Vx} \underseteq{eq:weight_mod}{\gets} 1/\Paren{\eta\data^{\Tag{mod}}\Paren{\Vx}+\varRON}$
		\commentalgo{Model-based confidence}

	\If{$\weight^{\Tag{rob}}\Paren{\Vx}<\percent{10}$ and $\Vx$ is isolated}
		\State $\weight\Paren{\Vx} \gets 0$		
			\commentalgo{Excluding isolated outliers}
			\label{lin:sig:th}
	\EndIf  

	\State $\V{\sigma} \underseteq{eq:sig_map}{\gets} \frac{\pmoon\corr\Brack{\Vweight\Paren{\Vdata - \Vdata^{\Tag{mod}}}}}{\sqrt{\Paren{\brack{\pmoon}^{2}\corr\Vweight}}}$
		\commentalgo{Significance map}
		\label{lin:sig:map}

	\State \Return $\V{\sigma}$
        
\end{algorithmic}
\end{algorithm}

Applying \refalg{alg:sig} leads to the map of \refsubfig{fig:detection}{a}. Compared with the resulting map of \refsubfig{fig:weight}{c}, the outliers and the object close vicinity are discarded but not the moons. Using this map of valid pixels in \refeq{eq:sig_map} leads to the significance map of \refsubfig{fig:detection}{c}, which is no more corrupted by the artifacts mentioned above. The three moon stand out of the noise, the faintest one (blue) being at the detection limit of $5\sigma$. The closest one (green), despite being in the concave side of the asteroid is unambiguously detected.

\section{RESULTS ON ON-SKY DATA}
\label{sec:results}

The method was tested on archival data on two asteroids of the main belt of the Solar system among the \unit{100}{\kilo\meter} class: \Kleopatra and \Elektra. The data were obtained respectively with the Near Infrared Camera 2 (NIRC2, ID U013N2, PI: de~Pater et al. 2008, Pk50 filter, $\lambda=\unit{1.6455}{\micro\meter}$, $\Delta\lambda=\unit{25}{\nano\meter}$) of the Keck II telescope and ZIMPOL (ID 199.C-0074, PI: Vernazza et al.\cite{Vernazza:21_Large_program} 2019, R filter, $\lambda=\unit{645.9}{\nano\meter}$, $\Delta\lambda=\unit{56.7}{\nano\meter}$), mounted on the Spectro Polarimetric High-contrast Exoplanet REsearcher\cite{Beuzit:19_SPHERE} (SPHERE) of the Very  Large Telescope (VLT) observatory, equipped with the SPHERE Adaptive eXtreme Optics system\cite{Fusco:16_SAXO} (SAXO). \Kleopatra has a dumbbell shape and is orbited by two known moons\cite{Ostro:00_Kleopatra, Descamps:11_Kleopatra_triple, Shepard:18_Kleopatra_radar, Marchis:21_Kleopatra}. \Elektra is surrounded by three moons\cite{Fuksa:23_Elektra_orbit}. Due to the lack of proper data reduction and data analysis tools\cite{Yang:16_Elektra_Minerva, Berdeu:22_Elektra}, its faintest companion was discovered only recently in archival data obtained in 2014 with the Integral Field Spectrograph of SPHERE\cite{Claudi:08_SPHERE_IFS} (IFS).

\begin{figure}[p!]
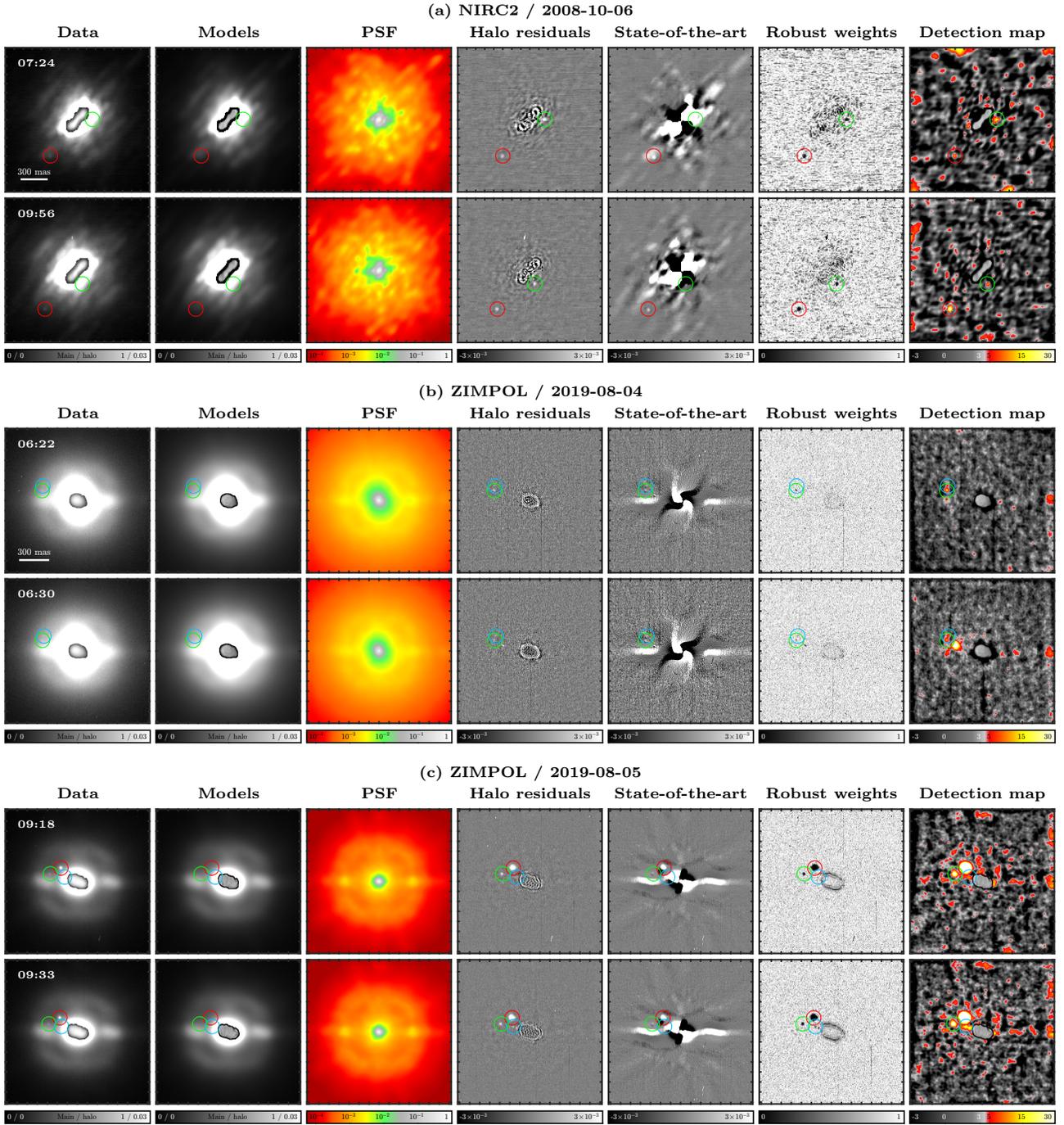
 % fig:on-sky
        \centering

        \newcommand{\PathFig}{Fig_comp_instru/}
        
        \newcommand{\subfigColor}{white}   
        
        % Line ratio
        \newcommand{\LineRatio}{0.975}    
        
        % Width of the text boxes        
        \newcommand{\widthTxt}{12pt}
        \newcommand{\widthFig}{\linewidth}
        
        % Vertical space between lines
        \newcommand{\spaceLine}{-0.1cm}
        
        % Font of the text in the figure
        \newcommand{\fontTxt}[1]{\textbf{\scriptsize #1}}
        
        % Defining column width command
        \newcommand{\ColumnWidth}
                {\dimexpr \LineRatio \widthFig / 7 \relax
                }
        \newcommand{\ColumnGap}{\hspace {\dimexpr \widthFig /7 - \LineRatio\widthFig /7 }}

        \newcommand{\inserLine}[3]{
                \subfigimg[width=\linewidth,pos=ul,font=\fontfig{\subfigColor}]{\tiny $\;$ #2}{0.0}{\PathFig #1_Data.pdf} &
                \subfigimg[width=\linewidth,pos=ul,font=\fontfig{\subfigColor}]{}{0.0}{\PathFig #1_Model.pdf} &
                \subfigimg[width=\linewidth,pos=ul,font=\fontfig{\subfigColor}]{}{0.0}{\PathFig #1_PSF.pdf} &
                \subfigimg[width=\linewidth,pos=ul,font=\fontfig{\subfigColor}]{}{0.0}{\PathFig #1_Res.pdf} &
                \subfigimg[width=\linewidth,pos=ul,font=\fontfig{\subfigColor}]{}{0.0}{\PathFig #1_Res_coro.pdf} &
                \subfigimg[width=\linewidth,pos=ul,font=\fontfig{\subfigColor}]{}{0.0}{\PathFig #1_Weight.pdf} &
                \subfigimg[width=\linewidth,pos=ul,font=\fontfig{\subfigColor}]{}{0.0}{\PathFig #1_Sig_BP.pdf}
                \\[#3]
                }
        
		\newcommand{\inserDoubleLine}[7]{
        		\multicolumn{7}{c}{\fontTxt{#1}}\\[-5pt]
            \fontTxt{Data}
            &
            \fontTxt{Models}
            &
            \fontTxt{PSF}
            &
            \fontTxt{Halo residuals}
            &
            \fontTxt{State-of-the-art}
            &
            \fontTxt{Robust weights}
            &
            \fontTxt{Detection map}
            \\
        		\inserLine{#2}{#4}{#6}
        		\inserLine{#3}{#5}{#7}
                \inserLine{Bar}{}{0pt} 
                }
                 
        % Figure table
        \begin{tabular}{
                @{}
                M{\ColumnWidth}
                @{\ColumnGap}
                M{\ColumnWidth}
                @{\ColumnGap}
                M{\ColumnWidth}
                @{\ColumnGap}
                M{\ColumnWidth}
                @{\ColumnGap}
                M{\ColumnWidth}
                @{\ColumnGap}
                M{\ColumnWidth}
                @{\ColumnGap}
                M{\ColumnWidth}
                @{}
                }

				\inserDoubleLine{(a) NIRC2 / 2008-10-06}{NIRC2_12003}{NIRC2_14003}{07:24}{09:56}{\spaceLine}{-6pt}
				\inserDoubleLine{(b) ZIMPOL / 2019-08-04}{ZIMPOL_2019-08-04_1}{ZIMPOL_2019-08-04_2}{06:22}{06:30}{\spaceLine}{-6pt}
				\inserDoubleLine{(c) ZIMPOL / 2019-08-05}{ZIMPOL_2019-08-05_1}{ZIMPOL_2019-08-05_2}{09:18}{09:33}{\spaceLine}{-6pt}         
        \end{tabular}        
        \caption{\label{fig:on-sky} Application of the method on different on-sky data. For each epoch, two frames at different timing (hh:mm) are presented along with: (i) the reduced blurred data $\Vdata$ and the saturated halo using a dual linear scale, (ii) the deconvolved object~$\tilde{\Vobj}$ and the resulting halo~$\Vdata^\Tag{mod}$ model using a dual linear scale, (iii) the deconvolved PSF~$\tilde{\Vpsf}$ with peak normalized to one, (iv) the fit residuals~$\Vdata - \Vdata^\Tag{mod}$, (v) the standard ``coronograph image'' obtained with state-of-the-art methods\cite{Pajuelo:18_Carry_coronagraph}, (vi) the equivalent robust weight map~$\weightrob$, and (vii) the detection map~$\V{\sigma}$. \refpans{a}: NIRC2 data on \Kleopatra. \refpans{b,c}: ZIMPOL data on \Elektra.}
\end{figure}

For each target, several frames are acquired at different epochs. To show the effectiveness of our method, we processed two individual frames of different epochs rather than the stacked observations. Results are gathered in \reffig{fig:on-sky}. For information, the ``coronagraph image'' rendered by state-of-the-art algorithm in asteroid study\cite{Pajuelo:18_Carry_coronagraph}, based on local median filtering, is also given.

Concerning the main object deconvolution, a sharp image with details on the body surface is retrieved for all datasets, even in bad seeing condition, see \refsubfig{fig:on-sky}{b}. The multiple images of \Kleopatra produced by the speckles of the complex PSF are correctly refocused in a single main object. When it comes to the reconstructed PSFs, they look similar to reference PSFs obtained on star (see \citet{Berdeu:24_deconv} for NIRC2 and \refsubfig{fig:overview}{a3} for ZIMPOL) and the expected features are present: (I) from to the low number of actuators of its AO system, many speckles are visible in the NIRC2 PSF and the hexagonal shape of its first Airy ring, due to the hexagonal shape of the telescope primary mirror segments is clearly marked, and (II) the AO cutoff frequency of the extreme AO of ZIMPOL is clear as well as its main speckles. The consistency in each epoch between the two frames of the details on the object's surface and of features of the PSFs supports the fact that the method is robust to turbulence conditions and that the retrieved features are real.

Most of the halo is removed from the residual maps. Nonetheless, structured signals are still visible on the primary edges, implying a too strong regularizing and a biased reconstruction. This is a necessary trade-off to avoid noise propagation and ensure the split between the PSF and the object in the blind deconvolution. Nonetheless, it must be noted that the amplitude of the residuals is hundred times smaller than the object dynamics. In all frames, the moons are clearly visible, even in the unfavorable dataset of 2019-08-04 which is noisy with a strong turbulence\footnote{We note here that the brightest moon of \Elektra is outside the field of view.}. Their counterpart can also be identified in the robust weight map. This is also true for the faintest moon (blue), very close the primary in the 2019-08-05 dataset, at distances non achievable with standard techniques.

Looking at the residuals provided by the state-of-the-art approach further highlights the gain achieved with the proposed method. The highly structured halo of the NIRC2 data is not properly removed and the closest moon (green) is invisible in the two frames. The furthest moon (red) is detectable only due to its punctual nature compared to the residual features. The halo removal seems a bit more efficient on ZIMPOL data except closed to the bright lobes produced by the strong speckles on the AO cutoff frequency ring. In the 2019-08-04 frames, the two faintest moon are barely visible compared to the noise level and only because they are luckily not in a region with residual artifacts. Too close from the primary, the faintest moon stays hidden by these residuals in the 2019-08-05 dataset. As a final remark, we recall here that this ``coronograph images'' are obtained using the full stack of the observed frames\footnote{Hence the same images for the ZIMPOL datasets of 2019-08-04 and 05. For the NIRC2 data, the presented frames come from two different observations performed with several hours of delay during the same night.} whereas our approach is presented on individual frames, allowing the study of the small orbital arc for each epoch where the stacked approach blurs this information.

Looking at the detection maps, except for a strong outlier visible on the second frame of the ZIMPOL 2019-08-04 dataset, most of the dead pixels and other sensor artifacts seem correctly removed. All the moons are properly detected above the $5\sigma$ threshold. Despite this good performance the results are less convincing than with the simulations: the maps are polluted with lots of false detection areas which seem to fit some large scale structures still present in the residuals. This could indicate that the smoothness regularization enforced on the PSF deconvolution is a bit too strong and that the highest spatial frequencies of the halo are not properly removed, leaving these structures in the residuals. The fact that lots of area are above the $5\sigma$ detection threshold also suggests that the weight factor $\weight$ in \refeq{eq:sig_map} is not properly scaled to really provide meaningful signal over noise ratios. Looking at \refeq{eq:weight_mod}, this means that the noise terms $\Paren{\eta, \varRON}$ are not properly calibrated. As discussed in \citet{Berdeu:24_deconv}, these terms are empirically (and pragmatically) estimated directly in the data of interested. This wrong scaling calls for a better refinement of the noise model along with the blind deconvolution iterations, as proposed in this reference\footnote{See also the documentation of the Epifluorescence DEconvolution MICroscopy plug-in\cite{Soulez:12} (EpiDEMIC) at \href{https://icy.bioimageanalysis.org/plugin/epidemic/}{https:// icy.bioimageanalysis.org/plugin/epidemic/}.}.

\section{Conclusions and perspectives}

In this work, we presented the performances of a blind and robust deconvolution algorithm, that jointly retrieves the asteroid image and the associated AO-PSF. Without any prior on the instrument nor the object shape (except that it has sharp edges), we showed with real data that it works with different instruments and in different turbulence conditions. It recovers details on the objects' surface and, efficiently retrieves the features of the PSFs specific to each instrument. This provides a physical and realistic model of the bright halo corrupting the data that can be subtracted to enhance the signal of potential close companions.

We also introduced a detection method to find these companions hidden in the residual noise.
We have shown that most of the outliers such as defective pixels or cosmic rays, are successfully robustly identified and discarded on-the-fly by the proposed algorithm. Nonetheless, some of them are missed. Added with some artifactual correlations induced by the structured halo removal, this degrades the quality of the significance maps. As already discussed above, it seems necessary to further refine the noise model to better estimate the signal of noise ratios. In addition, it would be possible to make the method more robust and also more sensitive to faint moons by combining multiple frames from a given epoch\cite{Thiebaut:16_detection}, or even from multiple epochs\cite{Dallant:22_PACOME}. Indeed, doing so provides another opportunity to identify random outliers on one side, while improving moon signal over noise ratios on the other side. Compared with the standard problem of exoplanet detection, it comes nonetheless with the additional difficulty of the rapid motion of the moons around the main object\cite{Showalter:19_Neptune, Berdeu:22_Elektra}.

\newpage

\acknowledgments % equivalent to \section*{ACKNOWLEDGMENTS}       
 
This project has received funding from the European Un{\-}ion's Horizon 2020 research and innovation programme under grant agreement No 101004719. This project has received funding from the French ``Agence nationale de la recherche'' under grant agreement No ANR-21-CE31-0015 (DDisk). 
\\
All the ZIMPOL reductions are based on public data acquired for the ESO Large Programme ID 199.C-0074 (Vernazza et al. 2021, `Asteroids as tracers of Solar System formation: Probing the interior of primordial main belt asteroids') and available at \href{https://observations.lam.fr/astero/}{https://observations.lam.fr/astero/
} and obtained via observations made with ESO Telescopes at the Para{\-}nal Observatory.
\\
The NIRC2 reductions are based data provided by the Keck Observatory Archive (KOA), which is operated by the W. M. Keck Observatory and the NASA Exoplanet Science Institute (NExScI), under contract with the National Aeronautics and Space Administration.

% References
\bibliography{2024_SPIE_Berdeu-et-al} % bibliography data in 2024_SPIE_Berdeu-et-al.bib

\begin{thebibliography}{10}

\bibitem{Vernazza:21_Large_program}
{Vernazza}, P., {Ferrais}, M., {Jorda}, L., and \etal, ``{VLT/SPHERE imaging
  survey of the largest main-belt asteroids: Final results and synthesis},''
  {\em A\&A}~{\bf 654},  A56 (Oct. 2021).

\bibitem{Roddier:81}
Roddier, F., ``V the effects of atmospheric turbulence in optical astronomy,''
  in [{\em Progress in Optics}{\nolinebreak\hspace{0.1em}]},  Wolf, E., ed.,
  {\bf 19},  281 -- 376, Elsevier (1981).

\bibitem{Tyson:15_principles_of_AO}
Tyson, R.,  [{\em Principles of Adaptive Optics}{\nolinebreak\hspace{0.1em}]},
  CRC Press (2015).

\bibitem{Jovanovic:15_XAO}
{Jovanovic}, N., {Martinache}, F., {Guyon}, O., and \etal, ``{The Subaru
  Coronagraphic Extreme Adaptive Optics System: Enabling High-Contrast Imaging
  on Solar-System Scales},'' {\em Publ. Astron. Soc. Pac.}~{\bf 127}(955),  890
  (2015).

\bibitem{Beltramo:20_PSF_reconstruction_Review}
Beltramo-Martin, O., Ragland, S., Fétick, R., and \etal, ``{Review of PSF
  reconstruction methods and application to post-processing},'' {\em SPIE Conf.
  Ser.}~{\bf 11448},  22 -- 36 (2020).

\bibitem{Marchis:06_moon_detection}
{Marchis}, F., {Kaasalainen}, M., {Hom}, E.~F.~Y., {Berthier}, J., {Enriquez},
  J., {Hestroffer}, D., {Le Mignant}, D., and {de Pater}, I., ``{Shape, size
  and multiplicity of main-belt asteroids. I. Keck Adaptive Optics survey},''
  {\em Icarus}~{\bf 185},  39--63 (Nov. 2006).

\bibitem{Fetick:19_Vesta}
{F{\'e}tick}, R.~J., {Jorda}, L., {Vernazza}, P., and \etal, ``{Closing the gap
  between Earth-based and interplanetary mission observations: Vesta seen by
  VLT/SPHERE},'' {\em A\&A}~{\bf 623},  A6 (Mar. 2019).

\bibitem{Fetick:20_param_marignal}
F{\'{e}}tick, R., Mugnier, L., Fusco, T., and Neichel, B., ``Blind
  deconvolution in astronomy with adaptive optics: the parametric marginal
  approach,'' {\em Mon. Not. Roy. Astron. Soc.}~{\bf 496},  4209--4220 (July
  2020).

\bibitem{Lau:23_prior_AOPSF}
{Lau}, A., {F{\'e}tick}, R.~J.~L., and {Neichel}, e., ``{Improved prior for
  adaptive optics point spread function estimation from science images:
  Application for deconvolution},'' {\em A\&A}~{\bf 673},  A72 (May 2023).

\bibitem{Veran:97_PSF_AO_telemetry}
Véran, J.-P., Rigaut, F., Maître, H., and Rouan, D., ``Estimation of the
  adaptive optics long-exposure point-spread function using control loop
  data,'' {\em J. Opt. Soc. Am. A}~{\bf 14},  3057--3069 (Nov 1997).

\bibitem{Clenet:08_NACO_PSF_recons}
{Cl{\'e}net}, Y., {Lidman}, C., {Gendron}, E., and \etal, ``{Tests of the PSF
  reconstruction algorithm for NACO/VLT},'' {\em SPIE Conf. Ser.}~{\bf 7015},
  701529 (2008).

\bibitem{Mugnier:04_Mistral}
{Mugnier}, L.~M., {Fusco}, T., and {Conan}, J.-M., ``{MISTRAL: a myopic
  edge-preserving image restoration method, with application to astronomical
  adaptive-optics-corrected long-exposure images},'' {\em J. Opt. Soc. Am.
  A}~{\bf 21},  1841--1854 (Oct. 2004).

\bibitem{Stockham:75_blind_deconvolution}
Stockham, T., Cannon, T., and Ingebretsen, R., ``Blind deconvolution through
  digital signal processing,'' {\em Proc. IEEE}~{\bf 63}(4),  678--692 (1975).

\bibitem{Thiebaut:95_blind_deconvolution}
{Thi{\'e}baut}, E. and {Conan}, J.~M., ``{Strict a priori constraints for
  maximum-likelihood blind deconvolution},'' {\em J. Opt. Soc. Am. A}~{\bf 12},
   485--492 (Mar. 1995).

\bibitem{Soulez:12}
Soulez, F., Denis, L., Tourneur, Y., and Thiébaut, {\'E}., ``Blind
  deconvolution of 3d data in wide field fluorescence microscopy,'' {\em IEEE
  ISBI}~{\bf 0},  1735--1738 (2012).

\bibitem{Blanc:03_Marginal_Zernike}
{Blanc}, A., {Mugnier}, L.~M., and {Idier}, J., ``{Marginal estimation of
  aberrations and image restoration by use of phase diversity},'' {\em J. Opt.
  Soc. Am. A}~{\bf 20},  1035--1045 (June 2003).

\bibitem{Fetick:19_model_based_AOPSF}
{F{\'e}tick}, R.~J.~L., {Fusco}, T., {Neichel}, B., and \etal, ``{Physics-based
  model of the adaptive-optics-corrected point spread function. Applications to
  the SPHERE/ZIMPOL and MUSE instruments},'' {\em A\&A}~{\bf 628},  A99 (Aug.
  2019).

\bibitem{Showalter:06_Uranus}
{Showalter}, M.~R. and {Lissauer}, J.~J., ``{The Second Ring-Moon System of
  Uranus: Discovery and Dynamics},'' {\em Science}~{\bf 311},  973--977 (Feb.
  2006).

\bibitem{Assafin:08_digital_coronography}
{Assafin}, M., {Campos}, R.~P., {Vieira Martins}, R., and \etal,
  ``{Instrumental and digital coronagraphy for the observation of the Uranus
  satellites{\textquoteright} upcoming mutual events},'' {\em Planet. Space
  Sci.}~{\bf 56},  1882--1887 (Nov. 2008).

\bibitem{Pajuelo:18_Carry_coronagraph}
{Pajuelo}, M., {Carry}, B., {Vachier}, F., and \etal, ``{Physical, spectral,
  and dynamical properties of asteroid (107) Camilla and its satellites},''
  {\em Icarus}~{\bf 309},  134--161 (July 2018).

\bibitem{Yang:16_Elektra_Minerva}
Yang, B., Wahhaj, Z., Beauvalet, L., and \etal, ``{Extreme AO observations of
  two triple asteroid systems with SPHERE},'' {\em Astrophys. J.}~{\bf 820},
  L35 (mar 2016).

\bibitem{Berdeu:22_Elektra}
{Berdeu}, A., {Langlois}, M., and {Vachier}, F., ``{First observation of a
  quadruple asteroid. Detection of a third moon around (130) Elektra with
  SPHERE/IFS},'' {\em A\&A}~{\bf 658},  L4 (Feb. 2022).

\bibitem{Berdeu:24_deconv}
Berdeu, A., ``{Blind and robust estimation of adaptive optics point spread
  function and diffuse halo with sharp-edged objects. Application to asteroid
  deconvolution and moon enhancement}.'' A\&A (in press), doi:
  \href{https://doi.org/10.1051/0004-6361/202347636
  }{10.1051/0004-6361/202347636 } (2024).

\bibitem{Schmid:18_ZIMPOL}
{Schmid}, H.~M., {Bazzon}, A., {Roelfsema}, R., and \etal, ``{SPHERE/ZIMPOL
  high resolution polarimetric imager. I. System overview, PSF parameters,
  coronagraphy, and polarimetry},'' {\em A\&A}~{\bf 619},  A9 (Nov. 2018).

\bibitem{Rudin:92_TV}
Rudin, L.~I., Osher, S., and Fatemi, E., ``Nonlinear total variation based
  noise removal algorithms,'' {\em J. Phys. D}~{\bf 60},  259--268 (Nov. 1992).

\bibitem{Charbonnier:97_TV}
Charbonnier, P., Blanc-Feraud, L., Aubert, G., and Barlaud, M., ``Deterministic
  edge-preserving regularization in computed imaging,'' {\em Trans. Image
  Process.}~{\bf 6},  298--311 (Feb. 1997).

\bibitem{Yan:23_myopic_MCMC}
Yan, A., Mugnier, L.~M., Giovannelli, J.-F., and \etal, ``{Marginalized myopic
  deconvolution of adaptive optics corrected images using Markov chain Monte
  Carlo methods},'' {\em J. Astron. 115 Telescopes Instrum. Syst.}~{\bf 9}(4),
  048004 (2023).

\bibitem{Berdeu:20_PIC}
{Berdeu}, A., {Soulez}, F., {Denis}, L., and \etal, ``{PIC: a data reduction
  algorithm for integral field spectrographs. Application to the SPHERE
  instrument},'' {\em A\&A}~{\bf 635},  A90 (Mar. 2020).

\bibitem{Moffat:69}
{Moffat}, A.~F.~J., ``{A Theoretical Investigation of Focal Stellar Images in
  the Photographic Emulsion and Application to Photographic Photometry},'' {\em
  A\&A}~{\bf 3},  455 (Dec. 1969).

\bibitem{Zoubir:18_robust_stat}
Zoubir, A.~M., Koivunen, V., Ollila, E., and Muma, M.,  [{\em Robust Statistics
  for Signal Processing}{\nolinebreak\hspace{0.1em}]}, Cambridge University
  Press (Oct. 2018).

\bibitem{Flasseur:19_PhD}
Flasseur, O., {\em Object detection and characterization from faint signals in
  images : applications in astronomy and microscopy}, PhD thesis, École
  doctorale Sciences Ingénierie Santé (Saint-Etienne) (2019).
\newblock Thèse de doctorat dirigée par Fournier, Corinne et Denis, Loïc
  Image Lyon 2019 / 2019LYSES042.

\bibitem{Hogg:79}
Hogg, R.~V., ``Statistical robustness: One view of its use in applications
  today,'' {\em Am. Stat.}~{\bf 33}(3),  108--115 (1979).

\bibitem{Huber:96}
Huber, P.,  [{\em Robust Statistical Procedures: Second
  Edition}{\nolinebreak\hspace{0.1em}]}, CBMS-NSF Regional Conference Series in
  Applied Mathematics, Society for Industrial and Applied Mathematics (1996).

\bibitem{Holland:77}
Holland, P.~W. and Welsch, R.~E., ``Robust regression using iteratively
  reweighted least-squares,'' {\em Commun. Stat. Theory Methods}~{\bf 6}(9),
  813--827 (1977).

\bibitem{Sigl:16_nonlinear_IRLS}
Sigl, J., ``Nonlinear residual minimization by iteratively reweighted least
  squares,'' {\em Computational Optimization and Applications}~{\bf 64},
  755–792 (Feb. 2016).

\bibitem{Thiebaut:16_detection}
Thi{\'{e}}baut, {\'{E}}., Denis, L., Mugnier, L., and \etal, ``Fast and robust
  exo-planet detection in multi-spectral, multi-temporal data,'' {\em SPIE
  Conf. Ser.}~{\bf 990957} (2016).

\bibitem{Gonzalez:20_Matlab}
Gonzalez, R., Richard, E., and Steven, L.,  [{\em Digital Image Processing
  Using MATLAB}{\nolinebreak\hspace{0.1em}]}, Knoxville: Gatesmark Publishing,
  third~ed. (2020).

\bibitem{Beuzit:19_SPHERE}
Beuzit, J.-L., Vigan, A., Mouillet, D., and \etal, ``Sphere: the exoplanet
  imager for the very large telescope,'' {\em A\&A}~{\bf 631},  A155 (2019).

\bibitem{Fusco:16_SAXO}
{Fusco}, T., {Sauvage}, J.~F., {Mouillet}, D., and \etal, ``{SAXO, the SPHERE
  extreme AO system: on-sky final performance and future improvements},'' {\em
  SPIE Conf. Ser.}~{\bf 9909},  99090U (2016).

\bibitem{Ostro:00_Kleopatra}
{Ostro}, S.~J., {Hudson}, R.~S., {Nolan}, M.~C., and \etal, ``{Radar
  Observations of Asteroid 216 Kleopatra},'' {\em Science}~{\bf 288},  836--839
  (May 2000).

\bibitem{Descamps:11_Kleopatra_triple}
{Descamps}, P., {Marchis}, F., {Berthier}, J., and \etal, ``{Triplicity and
  physical characteristics of Asteroid (216) Kleopatra},'' {\em Icarus}~{\bf
  211},  1022--1033 (Feb. 2011).

\bibitem{Shepard:18_Kleopatra_radar}
{Shepard}, M.~K., {Timerson}, B., {Scheeres}, D.~J., and \etal, ``{A revised
  shape model of asteroid (216) Kleopatra},'' {\em Icarus}~{\bf 311},  197--209
  (Sept. 2018).

\bibitem{Marchis:21_Kleopatra}
{Marchis}, F., {Jorda}, L., {Vernazza}, P., and \etal, ``{(216) Kleopatra, a
  low density critically rotating M-type asteroid},'' {\em A\&A}~{\bf 653},
  A57 (Sept. 2021).

\bibitem{Fuksa:23_Elektra_orbit}
{Fuksa, M.}, {Brož, M.}, {Hanuš, J.}, and \etal, ``An advanced multipole
  model of the (130) elektra quadruple system,'' {\em A\&A}~{\bf 677},  A189
  (2023).

\bibitem{Claudi:08_SPHERE_IFS}
{Claudi}, R.~U., {Turatto}, M., {Gratton}, R.~G., and \etal, ``{SPHERE IFS: the
  spectro differential imager of the VLT for exoplanets search},'' {\em SPIE
  Conf. Ser.}~{\bf 7014},  70143E (2008).

\bibitem{Dallant:22_PACOME}
{Dallant}, J., {Langlois}, M., {Thi{\'e}baut}, {\'E}., and {Flasseur}, O.,
  ``{Optimal multi-epoch combination of direct imaging observations for
  improved exoplanet detection},'' {\em SPIE Conf. Ser.}~{\bf 12185},  1218537
  (2022).

\bibitem{Showalter:19_Neptune}
{Showalter}, M.~R., {de Pater}, I., {Lissauer}, J.~J., and {French}, R.~S.,
  ``{The seventh inner moon of Neptune},'' {\em Nature}~{\bf 566},  350--353
  (Feb. 2019).

\end{thebibliography}
\bibliographystyle{spiebib} % makes bibtex use spiebib.bst

\end{document}